\documentclass[fleqn,10pt]{wlscirep}
\usepackage[english]{babel}
\usepackage[utf8]{inputenc}

\begin{document}

\title{Nature of the Dirac gap modulation and surface magnetic interaction in axion antiferromagnetic topological insulator MnBi$_2$Te$_4$}

\author[1,*]{A.~M.~Shikin}
\author[1]{D.~A.~Estyunin}
\author[1]{I.~I.~Klimovskikh}
\author[1]{S.~O.~Filnov}
\author[2]{E.~F.~Schwier}
\author[2]{S.~Kumar}
\author[2]{K.~Myamoto}
\author[2]{T.~Okuda}
\author[3]{A.~Kimura}
\author[4]{K.~Kuroda}
\author[4]{K.~Yaji}
\author[4]{S.~Shin}
\author[5]{Y.~Takeda}
\author[5]{Y.~Saitoh}
\author[6,7]{Z. S.~Aliev}
\author[7]{N.~T.~Mamedov}
\author[7]{I.~R.~Amiraslanov}
\author[8]{M.~B.~Babanly}
\author[9,10]{M.~M.~Otrokov}
\author[1,11,12]{S.~V.~Eremeev}
\author[1,13]{E.~V.~Chulkov}

\affil[1]{Saint Petersburg State University, 198504 Saint Petersburg,  Russia}
\affil[2]{Hiroshima Synchrotron Radiation Center, Hiroshima University, Hiroshima, Japan}
\affil[3]{Department of Physical Sciences, Graduate School of Science, Hiroshima University, Hiroshima, Japan}
\affil[4]{ISSP, University of Tokyo, Kashiwa, Chiba 277-8581, Japan}
\affil[5]{Materials Sciences Research Center, Japan Atomic Energy Agency, Sayo, Hyogo 679-5148, Japan}
\affil[6]{Azerbaijan State Oil and Industry University, Baku, Azerbaijan}
\affil[7]{Institute of Physics, Azerbaijan National Academy of Sciences, Baku, Azerbaijan}
\affil[8]{Institute of Catalysis and Inorganic Chemistry, Azerbaijan National Academy of Sciences, Baku, Azerbaijan}
\affil[9]{Centro de F\'isica de Materiales (CFM-MPC), Centro Mixto CSIC-UPV/EHU, E-20018 Donostia-San Sebasti\'an, Basque Country, Spain}
\affil[10]{IKERBASQUE, Basque Foundation for Science, E-48011 Bilbao, Basque Country, Spain}
\affil[11]{Institute of Strength Physics and Materials Science, 634055 Tomsk, Russia}
\affil[12]{Tomsk State University, 634050 Tomsk, Russia}
\affil[13]{Donostia International Physics Center (DIPC), 20018 Donostia-San Sebasti\'an, Basque Country,Spain}

\affil[*]{ashikin@inbox.ru}

\date{\today} 

\begin{abstract}

Modification of the gap  at the Dirac point (DP) in antiferromagnetic (AFM) axion topological insulator MnBi$_2$Te$_4$ and its electronic and spin structure has been studied by angle- and spin-resolved photoemission spectroscopy (ARPES) under laser excitation with variation of temperature (9-35~K), light polarization and photon energy. We have distinguished both a large (62-67~meV) and a reduced (15-18~meV) gap at the DP in the ARPES dispersions, which remains open above the N\'eel temperature ($T_\mathrm{N}=24.5$~K). We propose that the gap above $T_\mathrm{N}$  remains open  due to short-range magnetic field generated by chiral spin fluctuations. Spin-resolved ARPES, XMCD and circular dichroism ARPES measurements show a surface ferromagnetic ordering for large-gap sample and significantly reduced effective magnetic moment for the reduced-gap sample. These effects can be associated with  a shift of the topological DC state  towards the second Mn layer due to structural defects and mechanical disturbance, where  it is influenced by a compensated effect of opposite magnetic moments.
\end{abstract}

\maketitle

\section*{Introduction}

Interplay between topology and magnetism plays a significant role in
generation of a number of exotic topological quantum effects such as
Quantum Anomalous Hall Effect (QAHE)
\cite{Qi2008,Yu2010,Chang2013,Chang2015} and topological
magnetoelectric effect \cite{Qi2008,Wang2015,Chen2010}. These
effects,  discovered  in magnetic topological insulators
(TIs), are very important for both fundamental science and future
technological applications, like dissipation-less topological
electronics and topological quantum computation. They are
accompanied by a predicted opening of a gap at the DP arising as a
result of the time reversal symmetry (TRS) breaking due to the
induced magnetic ordering. In turn, this magnetic gap and its
magnitude can be good indicators of the developed effects and their
modification under different conditions. At the same time, magnetic
TIs (see, for example,
\cite{Otrokov2019,Otrokov.prl2019,Zhang2019,Li2019,Gong2019,Lee2019,Aliev2019,Hao2019,Chen2019,Swatek2019,
Shikin2019,Vidal.prb2019}), which are currently  viewed as  the best
candidates for implementing these effects, can also be considered as
a very promising platform for realizing  other interesting and exotic
effects, such as magnetic monopole and axion field action and their
possible manipulation
\cite{Qi2009,Qi2011,Qi2008,Witten1979,Rosenberg2010,Nogueira2016}.
These effects could also affect the magnitude of the  DP  magnetic
gap in the DP and could be considered as an indicator of the associated
effects of fractionalization of effective electron charge  and magnetic
flux quanta
\cite{Swingle2011,Rosenberg2010,Nogueira2016,Nogueira2018}.

Recently, an intrinsic magnetically-ordered AFM TI with
MnBi$_2$Te$_4$  stoichiometry, which demonstrates a combination of
antiferromagnetism and non-trivial bulk topology, has been
successfully synthesized, and its electronic structure has been
theoretically and experimentally investigated
\cite{Otrokov2019,Otrokov.prl2019,Zhang2019,Li2019,Gong2019,Lee2019,Aliev2019}. This compound has layered crystal structure consisting of septuple
layers (SLs) with van der Waals (vdW) bonding between them. Each SL
contains Mn atom layer in the middle plain with ferromagnetic coupling
between Mn magnetic moments. An overall AFM coupling in the compound is formed
by the nearest neighboring Mn FM layers stacked
antiferromagnetically along the out-of-plane direction. As a result,
all magnetic moments are ordered and aligned with the $c$-axis, i.e.
perpendicular to the surface. The  nontrivial
topological surface state is formed  at MnBi$_2$Te$_4$(0001)  thanks to the inverted
Bi $p_z$ and Te $p_z$ bulk bands at the $\Gamma$-point due to strong
spin-orbit coupling, which is essentially the same as for
Bi$_2$Te$_3$ \cite{Otrokov2019,Zhang2019,Li2019,Gong2019}. Although,
due to AFM coupling, TRS, $\Theta$, is broken, there exist two
symmetries: $\mathrm{P}_2\Theta$ and $\mathrm{S}=\Theta
\mathrm{T}_{1/2}$ (where $\mathrm{P}_2$ is an inversion operation
centered between neighboring layers and $\mathrm{T}_{1/2}$ is a
lattice translation), which preserve the topological invariant. I.e.
 the combination of these
two symmetries allows this compound to be AFM TI. Theoretical
calculations predict for AFM TI MnBi$_2$Te$_4$ a magnetically-driven
energy gap at the DP of  88~meV
\cite{Otrokov2019}.

At the same time, experimentally measured DP gap varies from 50 to 85 meV
depending on photon energy and the  measurement conditions, see, for
instance \cite{Otrokov2019,Zhang2019,Li2019,Gong2019,Lee2019}.
Moreover, it was found that the DC state remain  almost
temperature independent above and below the N\'eel temperature
($T_\mathrm{N}$), although certain temperature dependence of the
photoemission intensity and lineshape was also observed
\cite{Otrokov2019}.  Similar results, which show the gapped Dirac state at temperature above $T_\mathrm{N}$ were also reported for the Gd-doped
AFM TI \cite{Shikin2019,Shikin2020} and other kinds of
magnetically-doped TIs (see, for instance,
\cite{Chen2010,Xu2012,Shikin2018,Shikin2018a,Sanchez-Barriga2016,Filnov2019}).
At the same time, a series of works
\cite{Hao2019,Chen2019,Swatek2019,Li2019}  have
appeared in literature where a ''gapless''-like ARPES dispersion was
observed for MnBi$_2$Te$_4$. Although, for these measurements a
small gap of about 12-13.5 meV at the DP can be also distinguished,
which also remains open above  $T_\mathrm{N}$ \cite{Hao2019,Li2019}.
The reduced gap for these samples was explained as being
related to the difference between the bulk and surface magnetic
orders, although the reduced gap also showed a weak dependence on
temperature. The significant difference in the gap width
observed for different samples of MnBi$_2$Te$_4$ and its weak
dependence on the long-range magnetic order transition
 still have not been explained.

In the current work we have carried out a detailed analysis of the
Dirac gap  in MnBi$_2$Te$_4$ measured by Spin and Angle-resolved Photoemission
(ARPES) above and below $T_\mathrm{N}$ (24.5~K) with
variation of the photon energy (using laser and synchrotron
radiation (LR and SR)) with different light polarization. We show that the gap at the DP  in
this compound is slightly reduced, but remains open above
$T_\mathrm{N}$. We 
attribute it to the chiral-like spin fluctuations induced
above $T_\mathrm{N}$, which can be considered as a local emergent
magnetic field preserving the gap. We present and compare
the results of the ARPES measurement for different kinds of the
MnBi$_2$Te$_4$ samples (or different surface areas for one sample)
demonstrating a large (62-67 meV) and a reduced (15-18 meV) gap at
the DP. Both kinds of the samples are characterized by the same
perfect X-ray diffraction \cite{Aliev2019}.  We have studied  these samples by
XMCD, CD and spin-resolved ARPES and have showed a possibility
of difference in the formed magnetic moment, which affects the DC
state, in the case of the samples with different gap
at the DP. These effects can be related to a shift
of the topological DC state  towards the second 
septuple layer block, due to structural effect as revealed by our
\emph{ab-initio} simulations,  where it experiences the opposite
magnetic moments of the second Mn layer.  Thus, this shift leads to
a decrease in the effective magnetic moment, which  acts on the DC
state. Finally, the possibility of many-body fractionation
effects arising under coupling of the spin fluctuations of opposite
chirality generated in the nearest surface magnetic layers is
discussed.

\section*{Results}

\noindent
\textbf{ARPES dispersion maps}

\begin{figure*}
\centering
\includegraphics[width=0.8\textwidth]{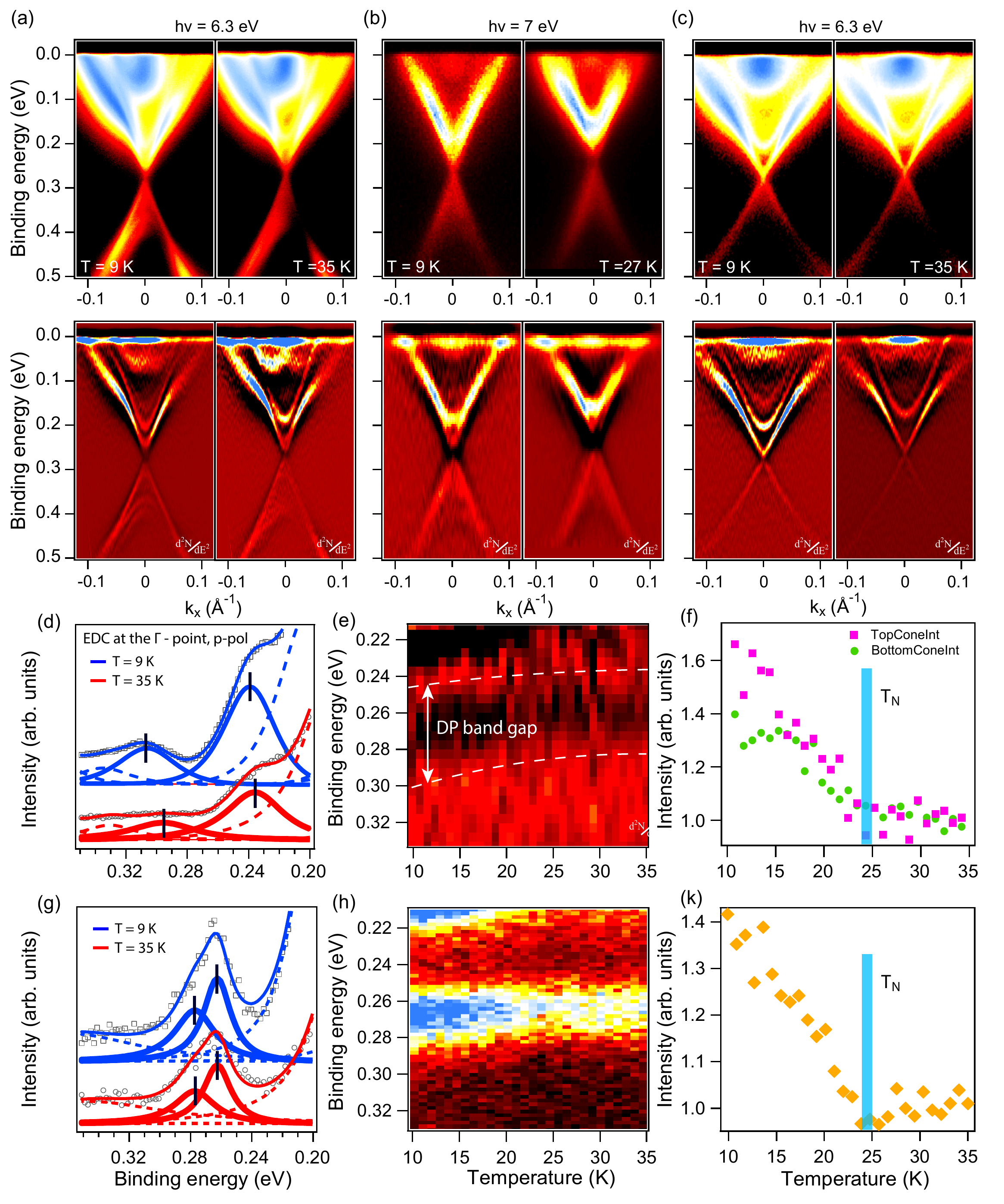}
\caption{(a,b), upper line - ARPES dispersion maps measured for MnBi$_2$Te$_4$ at different photon energies 6.3~eV and 7.0~eV using p-polarized laser radiation below (9~K) and above $T_\mathrm{N}$ (35~K and  27~K) – for sample with large gap. (a,b), second line – the same in the $d^2N/dE^2$ presentation. (c) – the same as in panels (a,b) measured for sample with reduced gap. (d) - EDCs measured for sample with large directly in the region of the DP at the $\Gamma$-point ($k_\parallel$=0) at temperatures of 9~K and 35~K (for the experiment presented in (a)), blue and red  curves, respectively, with presentation of deconvolution on spectral components. (e) - the variation of the Dirac gap value under permanent growth of temperature between 9~K and 35~K measured directly at the DP at $k_\parallel$=0 and (f) – the corresponding variation of the intensity of the DC state at the DP with temperature with presentation of the contributions of the upper and lower DC state peak intencities – for sample with large gap. (g,h,k) – the same as in panels (d,e,f) measured for sample with reduced gap.}
 \label{ARPES1}
\end{figure*}

Figs.~\ref{ARPES1}~(a and b, upper row) show a comparison between the ARPES dispersion maps measured for MnBi$_2$Te$_4$ below and above $T_\mathrm{N}$ ($T_\mathrm{N}$ = 24.5~K \cite{Otrokov2019,Aliev2019}) using $p$-polarized laser radiation with photon energies of 6.3~eV and 7.0~eV. (This kind of samples or surface areas hereafter we will call as samples with a large gap at the DP). Fig.~\ref{ARPES1}~(c) shows the corresponding ARPES dispersion maps for the second sample with reduced gap at the DP, see details in the text below. In the bottom row the corresponding ARPES maps in the $d^2N/dE^2$ form are shown for better visualization of the spectra, including  above $T_\mathrm{N}$. For the measurements presented in panels (a) and (c) the $\mu$-Laser ARPES system ($h\nu$=6.3eV) at the Hiroshima Synchrotron Radiation Center \cite{Iwasawa2017} with improved angle, energy resolution and spatial resolutions was used. The ARPES dispersion map presented in panel (b) were obtained during the spin-resolved ARPES experiments (see below) using LR with a photon energy of 7~eV \cite{Yaji2019}. From the dispersion maps at 9~K presented in panels (a) and (c)  the positions of the edges of the conduction and valence band (CB and VB) states  can be estimated as lying at Binding Energies (BEs) of about 0.22~eV and 0.36~eV, respectively. At temperature of 35~K the CB and VB edges are located approximately at BEs of 0.19 and 0.38~eV. This difference is related to the exchange splitting of the bulk states  below $T_\mathrm{N}$ (at 9~K) that is followed by decrease of a total fundamental gap  (see Ref.~\cite{Estyunin2020} for more details and analysis of these states and their splitting).  The ARPES dispersion maps presented in panel (a) demonstrate an  energy gap at  DP both below and above $T_\mathrm{N}$. 

Fig.~\ref{ARPES1} (d) shows the corresponding Energy Distribution Curves (EDCs) measured at the $\Gamma$-point ($k_\|$=0)  at $T$=9 and 35~K (blue and red curves, respectively) using $p$-polarized laser radiation of 6.3~eV. One can clearly see a significant dip in photoemission intensity at 9~K in the region of the gap between the maxima at the edges of the gap (marked by black vertical lines).  At the same time the dip and the edge gap  peaks remain visible above $T_\mathrm{N}$, too. It means that the gap size can be determined both below and above $T_\mathrm{N}$. The fitting allows to estimate the value of the gap at the DP (or the splitting between the maxima of the upper and lower DC states) at 9~K on the level of about 67~meV (62~meV for $s$-polarization). At 35~K this value is slightly decreasing till 59~meV (53~meV for $s$-polarization). Taken into account the width of the measured spectral lines the uncertainty in the measured gap value can be estimated as  $\pm$5~meV. Moreover, the estimation using EDCs measured at temperatures above $T_\mathrm{N}$ is made more difficult due to reduced intensities of the edge gap maxima despite the high resolution.

 To investigate the evolution of the DP gap as a function of temperature, we  performed a set of  measurements between 9 and 35~K, crossing $T_\mathrm{N}$. In  Fig.~\ref{ARPES1} (e) a direct variation of the intensity of edges of gap is presented where white dotted lines show maxima of intensities. The gap edges are located at the BEs of about 0.245~eV and 0.3~eV at 9~K and  they smoothly shift towards lower BEs with increased temperature.  As can be seen,  the Dirac gap indeed remains open above $T_\mathrm{N}$, however a  small reduction in its size is observed at temperature above 20-25~K. Additionally, Fig.~\ref{ARPES1} (f) demonstrates the corresponding variation of the spectral weight of gap edges with  temperature. The vertical blue line indicates $T_\mathrm{N}$ taken from the magnetometry (SQUID) measurements \cite{Otrokov2019,Estyunin2020}. One can clearly see that the intensity of the DC states decreases continuously with temperature approaching $T_\mathrm{N}$. At higher temperatures it remains at an approximately constant level. Thus, the presented results confirm the opening of a large gap at the DP with a value of about 62-67~meV as previously published \cite{Otrokov2019}.

At the same time, there are samples of the second kind, also measured under laser photoexcitation, which demonstrate fundamentally different  spectrum (so-called ``gapless''-like dispersion). Fig.~\ref{ARPES1}~(c) shows the  ARPES dispersion maps  for this kind of samples, measured at temperatures both below and above  $T_\mathrm{N}$ (9~K and 35~K). In Fig.~\ref{ARPES1}~(g) the corresponding EDCs measured at the $\Gamma$-point  are also presented.  At the first glance the ARPES dispersions presented in panel (c) demonstrate  a ``gapless''-like character.  Nevertheless, the EDCs presented in panel (g) allow to distinguish a small energy gap both at temperatures of 9~K and 35~K (shown by red and blue curves, respectively). Decomposition on the corresponding spectral components (also shown by red and blue lines below  the ``as-measured'' EDCs)  allows us to estimate the  the gap width  as about of 15-18~meV, both below and above $T_\mathrm{N}$. This value correlates with the gap  estimated in Ref.~\cite{Li2019}. While the exact gap size determination is subject to broadening effects, the presence of large and reduced gaps in MnBi$_2$Te$_4$ can be confirmed.

Fig.~\ref{ARPES1}~(h) directly demonstrates the behavior of the gap under crossing $T_\mathrm{N}$ using the PE intensity profile changes under permanent growth of the temperature between 9~K and 35~K. All visible variations of the line width are related to the intensity depleting. In reality, no significant variation of the averaged line width was distinguished. At the same time, as shown in Fig.~\ref{ARPES1}~(k), the intensity of the DC state at the DP reduces when $T_\mathrm{N}$ crosses, which also amounts to 25~K as for the sample with the large gap (panels (d)-(f)). This decrease in the DC state intensity correlates with the collapse of the exchange splitting of the CB edge states (see the full-range dispersion maps in Fig.~\ref{ARPES1}~(a) at $T_\mathrm{N}$. It is interesting that the  exchange splitting for the edge CB states for samples presented in Fig.~\ref{ARPES1} (a) and Fig.~\ref{ARPES1} (c) below $T_\mathrm{N}$ has almost the same value. This indicates the  identity of the bulk AFM ordering for the samples showing DC gaps  of 62-67~meV (Fig.~\ref{ARPES1} (a)) and 15-18~meV (Fig.~\ref{ARPES1}~(c)).

In Refs.~\cite{Sanchez-Barriga2016,Black-Schaffer2015,Xu2017} it was proposed that the gap at the DP, even in magnetically-doped TIs can be developed and remain open due to the avoided band crossing under hybridization of the DC states near the DP with the impurity levels of magnetic atoms, when they  have binding energies inside the bulk band gap near the DP.
To confirm a magnetic-derived origin of the gap opening (i.e. that the gap  at the DP above $T_\mathrm{N}$ in MnBi$_2$Te$_4$ is not related to the non-magnetic  hybridization mechanism based on avoided-crossing effects in formation of electronic structure proposed in Refs.~\cite{Sanchez-Barriga2016,Black-Schaffer2015,Xu2017}) we performed resonant PE measurements. The resonant PE experiment allows to enhance the contribution of the Mn-derived states in PE spectra and to indicate the exact energy range for localization of these states.

\noindent
\textbf{Resonant photoemission measurements}

\begin{figure*}
\centering
\includegraphics[width=0.8\textwidth]{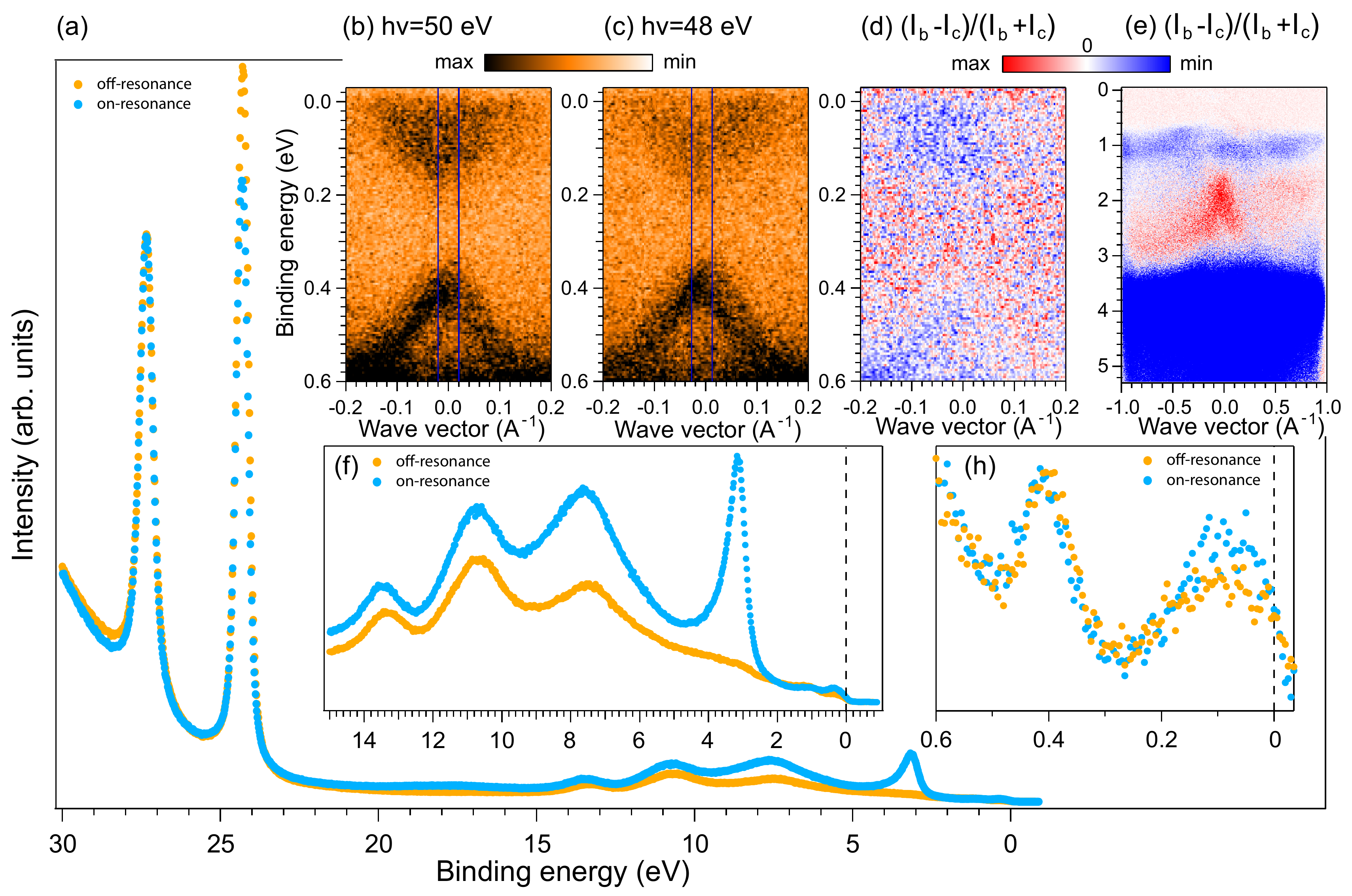}H
\caption{Main panel (a) - resonant Mn(3p-3d) photoexcitation PE spectra measured on-resonance ($h\nu$=50eV) and off-resonance ($h\nu$=48eV) which demonstrates the on-resonance increase of the intensity of the Mn(d)-derived states (blue curve in comparison with red one), see panel (f) in more details in the region of BEs till 15~eV. Panel (h) shows in details the region of the DC and CB states. The upper panels (b,c) – the on-resonance and off-resonance ARPES dispersion maps measured in the region of the fundamental gap between the CB and VB states (where the DP is located with BE of 0.28~eV). (d) – the difference between the on- and off resonance ARPES dispersion maps. (e) – the same as shown in panel (d), only till the higher BEs.}
 \label{PE}
\end{figure*}

According to theoretical calculations (see, for instance, \cite{Otrokov2019,Zhang2019,Jahangirli2019}),  in MnBi$_2$Te$_4$ the Mn 3$d$ states are located very far from the topological surface state. To confirm this fact experimentally Fig.~\ref{PE} demonstrates the results of  PE resonant experiment for MnBi$_2$Te$_4$. Fig.~\ref{PE} (a) shows the PE spectra measured at the photon energies corresponding to on-resonance ($h\nu$=50~eV) and off-resonance ($h\nu$=48~eV) conditions  for the Mn(3$p$-3$d$) photoexcitation threshold. Comparison between these spectra demonstrates the on-resonance increase of the intensity of the Mn($d$)-derived states. One can clearly see a resonant increase of the Mn 3$d$ states with BE of about 3.2~eV  (see blue and orange curves). At the same time, no Mn $d$ derived intensity increase is visible in the region of the fundamental energy gap.  It is worth noting that  the samples, which demonstrated  large and small gap during the laser ARPES measurements both show a large gap under measurements with use of synchrotron radiation with photon energy 48 and 50 eV  due to matrix element effects and bulk dispersion \cite{Hao2019}.  Panel (d) shows the difference between the on- and off-resonance ARPES dispersion maps demonstrating the absence of any notable Mn $d$ states in the region of the DC state (0.28~eV in correspondence with Fig.~\ref{ARPES1}). This is clearly opposed to the strong spectral intensity  in Mn-doped Bi$_2$Te$_3$ \cite{Ruzicka2015} and shows the unlikeliness of an avoided-crossing-scenario in MnBi$_2$Te$_4$. Panel (e) demonstrates the same difference between the on- and off-resonance in ARPES dispersions as in panel (d), but focusing on higher BEs. The bottom  panels (f,h) demonstrate the difference between the on- and off-resonance PE spectra measured at the $\Gamma$-point. Panels (e,h) show that some weak Mn-derived state character can be also distinguished in the region of the VB states with BE of about 1~eV and in the region of the CB states which can be related to the formation of the Mn($d$)-Te($p$) hybrid states \cite{Jahangirli2019}. However, these states do not significantly affect opening the gap at the DP in framework of a possible avoided-crossing hybridization mechanism. All measured samples demonstrate analogous behavior during the resonant photoemission experiment with a small variation of the intensity of the Mn-derived peak with BE of 3.2~eV.

\noindent
\textbf{In-plane and out-of-plane spin texture of the gapped Dirac  state}

\begin{figure*}
\centering
\includegraphics[width=0.8\textwidth]{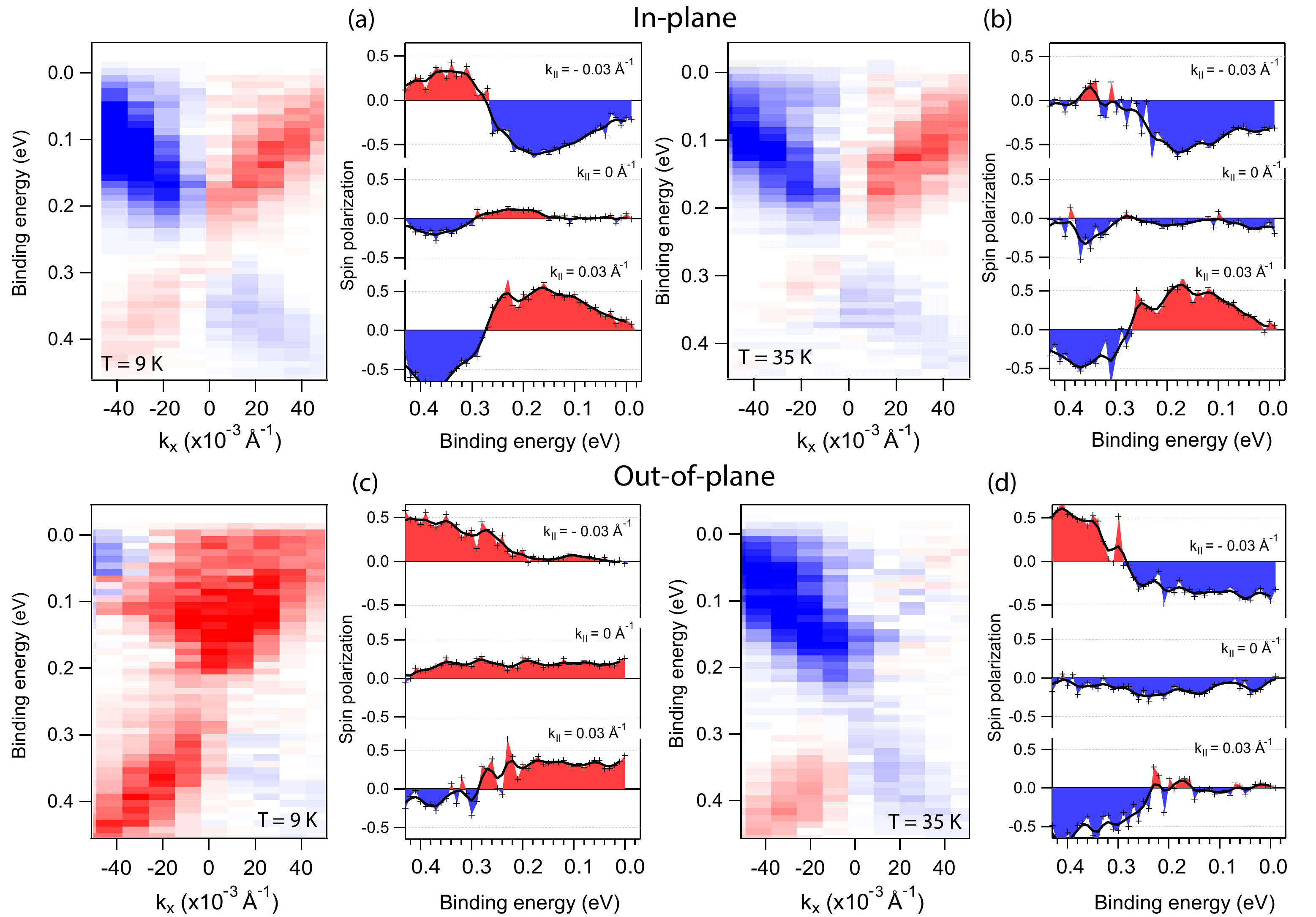}
\caption{(a,b), left insets,- in-plane spin-resolved ARPES dispersion maps measured for MnBi$_2$Te$_4$ using s-polarized laser radiation ($h\nu$ = 7~eV) at a temperature of 9~K and 35~K, respectively. Right insets – the corresponding in-plane polarizations measured at the $\Gamma$-point and symmetrically relative to it. (c,d) – the same as shown in panels (a,b), only with presentation of the out-of-plane spin-resolved ARPES dispersion maps and corresponding out-of-plane polarization.}
 \label{spin}
\end{figure*}

To analyze the spin structure of the DC state formed in MnBi$_2$Te$_4$ at temperatures below and above the $T_\mathrm{N}$ we have measured high resolution spin-resolved ARPES dispersion maps for the in-plane and out-of-plane spin polarization. The  spin-resolved ARPES dispersion maps measured along the $\Gamma$M direction with $s$-polarized LR are presented in Fig.~\ref{spin}. Fig.~\ref{spin}~(a and b, left panels) show the in-plane spin-resolved ARPES dispersion maps measured  using $s$-polarized LR ($h\nu$=7~eV) at temperature of 9~K and 35~K, respectively (below and above the $T_\mathrm{N}$). The right panels depict the binding energy-dependent variation of the in-plane spin polarization measured around the $\Gamma$-point.  The spin-resolved ARPES dispersion maps presented in panels (a,b) highlight the helical in-plane spin structure of the DC state and its topological character both below and above $T_\mathrm{N}$. One can clearly see the spin inversion with respect to $k_\|$ direction and between the upper and lower DC.  It is confirmed by the corresponding in-plane spin-polarization measurements at positive and negative $k_\|$ which are presented on right sides in the panels (a,b). Such spin polarization is observed for MnBi$_2$Te$_4$ both below and above $T_\mathrm{N}$. 

Figs.~\ref{spin}~(c and d, left panels) show the ARPES dispersion maps measured for the out-of-plane spin component using $s$-polarized LR at temperatures of 9~K and 35~K, respectively. The right panels show the corresponding variation of the out-of-plane spin polarization measured at different BEs at the $\Gamma$-point  and symmetrically at positive and negative $k_\|$. In opposite to the in-plane spin polarization measurements the presented spin-resolved dispersion maps for the out-of-plane polarization do not demonstrate a pronounced spin inversion between opposite momenta. While, for positive and negative momenta relative to the $\Gamma$-point the magnitude of the spin polarization is sharply changed at the energies corresponding to the DP position. This change takes place on the background of the constant-like contribution which can be related to the PE-induced spin polarization of the CB states and their hybridization with the Mn 3$d$ states. Such behavior of the out-of-plane polarization is observed both below and above $T_\mathrm{N}$, with some change of the polarization level with temperature. This observation may confirm that, above $T_\mathrm{N}$, the out-of-plane magnetization is also generated, despite the long-range magnetic order destroying. This is consistent with our assumption about preserving a magnetic-derived nature of the gap above $T_\mathrm{N}$, which is associated with the generation of an effective out-of-plane spin component in the developed spin fluctuations and corresponding spin texture.  Formation of such a combined out-of-plane spin structure correlates with the results of calculations of the PE-induced out-of-plane polarization for magnetically-doped TI in Ref.~\cite{Shikin2018a}, which arise due to the influence of the final state effects developed during PE process. Besides, the spin polarization of the Mn 3$d$  states located in the region of the CB states (see Fig.~\ref{PE}) also contributes to this spin-polarized background in the measured ARPES dispersion maps with dependence of both contributions on temperature and photon energy.

On the whole, the measured spin texture confirms that MnBi$_2$Te$_4$ is characterized by helical spin structure characteristic of TI. The out-of-plane polarization in the region of the DP observed in the EDC  and its inversion for the upper and lower DCs  testify to a magnetic origin of the gap  and its preservation above $T_\mathrm{N}$ as well. At the same time, the measured spin-resolved spectra are sensitive to the details of PE process and the contribution of the Mn-derived states varied with temperature.

\noindent
\textbf{Photon energy dependence of the out-of-plane polarization of the DC state}

\begin{figure*}
\centering
\includegraphics[width=0.8\textwidth]{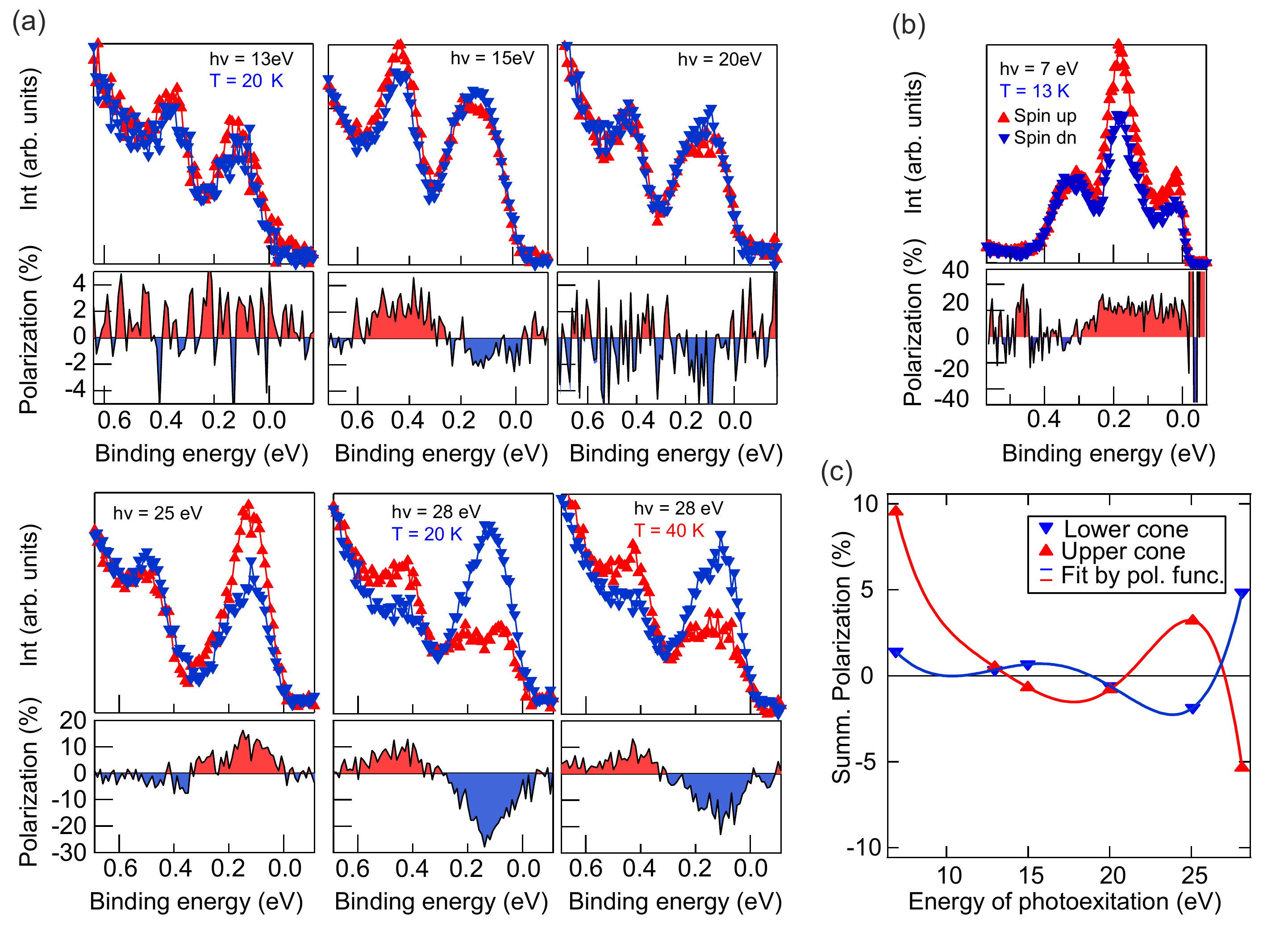}
\caption{(a) - Photon energy dependence ($h\nu$=13-28~eV) of the of out-of-plane spin polarization measured at the DP at $k_\parallel$=0 below ($T=20$~K) and above ($T=40$~K) the $T_\mathrm{N}$. Upper panels in each insets show the contribution of the spin-up and spin-down components in the measured EDCs. The lower panels demonstrate the resulting out-of-plane polarization. (b) - The same measured at photon energy of 7.0~eV at $T=13$~K (laser photoexcitation). (c) – Photon energy dependent variation of the out-of-plane polarization of the upper and lower DC state.}
 \label{Phdep}
\end{figure*}

To check the out-of-plane spin polarization inversion between the upper and lower branches of the gapped DC  and its possible variation on photon energy we have measured the spin-resolved PE spectra with the out-of-plane spin polarization at different photon energy under photoexcitation by synchrotron and laser radiation.

Figs.~\ref{Phdep} (a,b) show spin-resolved EDCs measured at different photon energy between 7~eV and 28~eV using $p$-polarized photons (both LR and SR) at temperatures below ($T=13$~K and 20~K) and above $T_\mathrm{N}$ ($T=40$~K). The bottom panels demonstrate the corresponding out-of-plane spin polarization for  the lower and upper DCs.  One can see that the sign and the magnitude of the out-of-plane polarization actually change depending on photon energy. The experimentally observed out-of-plane polarization changes sign in the region of photon energies between 28 and 25~eV (in the regions of photoexcitation of the Bi $5d_{3/2}$ and $5d_{5/2}$ states) and then oscillates with variation in the  photon energy in magnitude and sign. I.e. the variation of the out-of-plane polarization with photon energy has rather oscillating character (see panel (c)), as was calculated in Refs.~\cite{Shikin2018a,Zhu2013,Zhu2014} for TI. This observation can confirm that the PE-final state effects actually contribute to the measured spin structure of the PE-spectra in MnBi$_2$Te$_4$ as well. The contribution of the PE final state effect is changing with photon energy. At the same time, one can see that the spin-resolved spectrum measured at $h\nu$=7.0 eV (panel (b)) demonstrates an additional out-of-plane polarization contribution in the region of the CB states, almost independent on BE. We relate it  to the out-of-plane polarization of the CB states generated by PE which is enhanced at the photon energies between 6 and 15~eV due to their hybridization with the Mn 3$d$ states \cite{Jahangirli2019}. This enhancement is well visible in Fig.~\ref{PE} (f).

Thus, the presented spin-resolved spectra demonstrate a pronounced inversion of the out-of-plane spin polarization between the upper and lower gapped DC, while the value and the sign oscillate with photon energy in PE spectra  due to the developed final state effects. The observed out-of-plane spin polarization confirms a magnetic-derived nature of the gap  at the DP. The spin-resolved spectra taken at $h\nu$=28~eV at temperatures of 20 and 40~K demonstrate a slight decrease in the DC spin polarization above $T_\mathrm{N}$. The fact that out-of-plane spin polarization remains above $T_\mathrm{N}$ can support the idea of a magnetic-derived origin of the gap including above $T_\mathrm{N}$. Unfortunately, the use a large light spot during the SR experiment does not allow us to distinguish the samples (or surface areas) characterized by large and reduced gaps.

To analyze the difference in the measured gap at the DP for samples with a large and a small gaps (62-67 meV and 15-18 meV) and to try to find the reasons for this difference, we carry out below a direct comparison between these two kinds of samples by measuring the detailed ARPES dispersion maps at different polarization of LR and different temperatures.

\noindent
\textbf{Large and small gaps at different polarizations of LR and temperature}

\begin{figure*}
\centering
\includegraphics[width=0.8\textwidth]{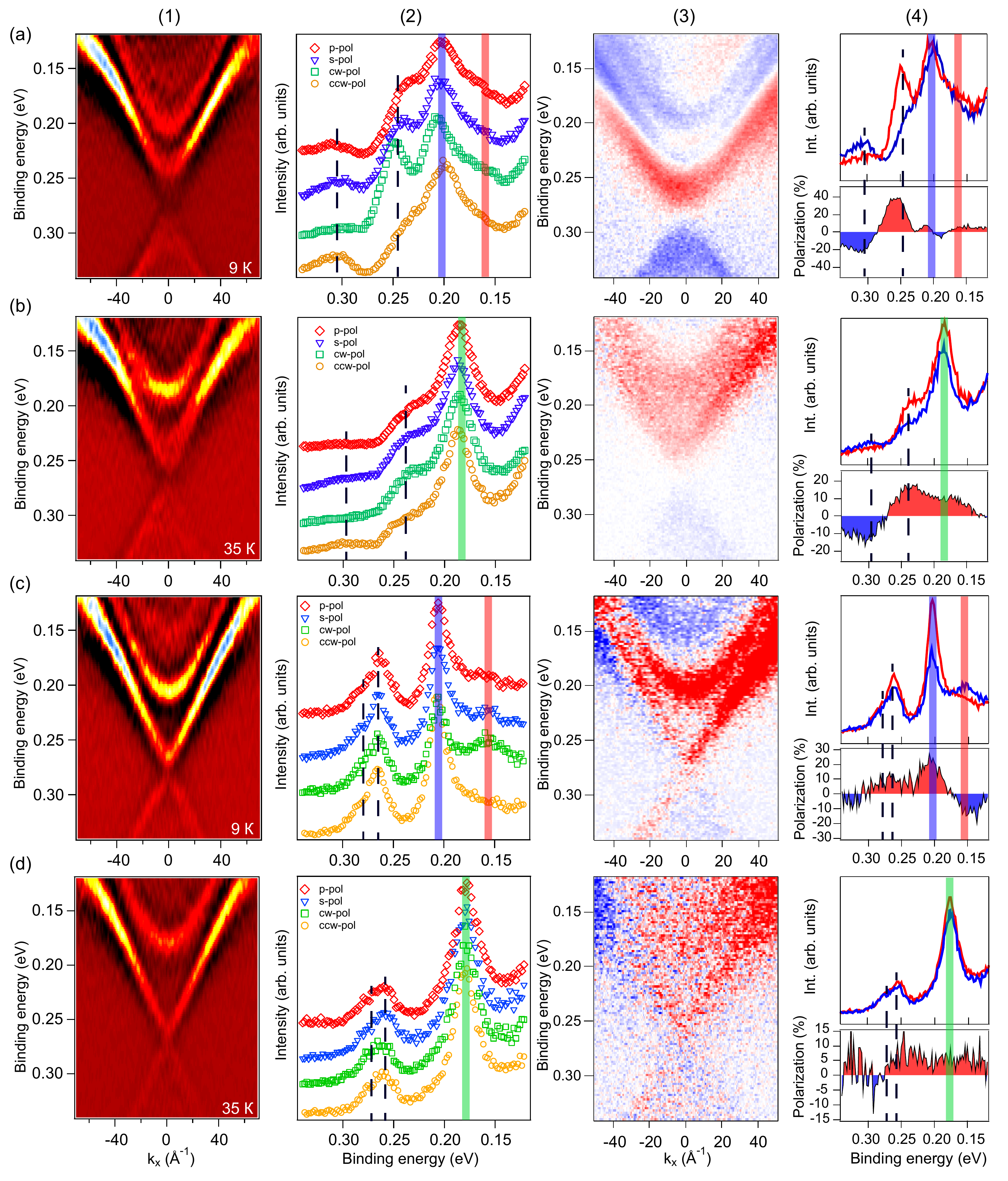}
\caption{(a1,b1,c1,d1) - the ARPES dispersion maps measured for the cases of a large (a1,b1) and reduced (c1,d1) gap at the DP at temperatures of 9~K (a1,c1) and 35~K (b1,d1) using p-polarized LR (with $h\nu$=6.3~eV). The spectra are shown in the region close to the DP in the $d^2N/dE^2$ presentation for better visualisation of the Dirac gap. (a2,b2,c2,d2) - the corresponding EDCs measured at the DP at $k_\parallel$=0 using different polarization of LR at temperature 9~K and 35~K, respectively, in the energy region close to the DP. (a3,b3,c3,d3) – the CD ARPES dispersion maps obtained by subtraction of the PE signal of opposite circular polarization. (a4,b4,c4,d4) – a comparison between the corresponding CD EDCs in the region close to the DP (upper panels) with presentation of the subtracted PE signal measured at opposite circular polarizations (bottom panels).}
 \label{ARPES2}
\end{figure*}

In Fig.~\ref{ARPES2} we compare the ARPES dispersion maps for samples with large (panels a,b) and reduced (panels c,d) gap, measured at different polarization of LR ($p$-, $s$-, and opposite circular ones) and different temperatures (9~K and 35~K). In Fig.~\ref{ARPES2} (panels a1,b1,c1,d1)  the corresponding ARPES dispersion maps in the $d^2N/dE^2$ form which were measured using $p$-polarized LR at $h\nu$=6.3~eV are presented. Figs.~\ref{ARPES2} (a2,b2,c2,d2) show the corresponding EDCs measured at the $\Gamma$-point  at different polarization of LR at temperatures of 9~K and 35~K, respectively. The maxima of intensity, corresponding to the edges of the DC gap  are marked by vertical black dashed lines. One can see that for the sample with  large gap (panels a,b) a significant redistribution of the detailed structure of the DC state at different polarization of LR takes place leading to an apparent gap modification. Most important  observation is a pronounced modification of the intensity of the upper and lower DC states at opposite circular polarizations. For positive circular polarisation the intensity of the upper DC  is enhanced. In contrast, for negative circular polarization an enhancement of the lower DC intensity  is observed. For temperature of 35~K the intensity of the DC state is reduced. However, the gap at the DC remains visible for all polarizations of LR. The tendency of the redistribution between the upper and lower DC states under circular polarization of LR remains the same as at 9~K. Panels a3, b3 and a4,b4 show this redistribution  in  details. In panels a3, b3 the circular dichroism (CD) ARPES dispersion maps are shown, which were derived by subtracting the spectra measured at opposite circular LR polarizations. Figs.~\ref{ARPES2} (a4 and b4, upper panels) compare the EDCs measured at opposite circular polarizations in the region close to the DC gap. The lower panels demonstrate the corresponding CD signal obtained by subtraction of the EDCs presented in upper panels with normalization on their sum.

The change in the sign of the resulting subtracted spectra correlates to some extent with the change in the spin polarization of the gapped DC state (see for Ref.~\cite{Kuch2001}). To confirm such a correlation, the spectra are presented in the region included also the Te-derived states at the edge of CB located in the energy region 0.14-0.22 eV, which are characterized by exchange splitting affected by the developed surface magnetic ordering (see for comparison Ref.~\cite{Estyunin2020}). These exchange split states are marked in the presented EDCs by blue and red vertical stripes for temperatures below $T_\mathrm{N}$ and by vertical green stripes above $T_\mathrm{N}$. Below $T_\mathrm{N}$, these states collapse in energy. We have to note that this energy region includes also the spin-polarized Mn-derived states. Fig.~\ref{PE} confirms their localization in this energy region.

For the case of  large gap at the DP at temperature below $T_\mathrm{N}$ (panels a3,a4), the surface topological DC state is mainly located in the first surface SL and, therefore, is affected by the first Mn-layer FM ordering. As a result, the CD-signal demonstrates a pronounced inversion of the CD-signal between the upper and lower DC states. At the same time, the opposite spin-polarization of the Mn-derived states and the Te-derived exchange-split states can lead to depletion of the inverted spin polarization in the resulting CD signal in this energy region (see panel a4).

Above $T_\mathrm{N}$, when the long-range magnetic ordering is destroyed, the energy splitting between the edge CB states collapses, and no sign inversion is observed for these states in the CD signal (panel b4). At the same time, the sign inversion in the CD signal between the upper and lower DC states remains visible above $T_\mathrm{N}$. The same inversion can be distinguished in the ARPES dispersion map (panel b3). This may indicate the preservation of a magnetic-derived origin of the gap  above $T_\mathrm{N}$, at least of a short-range character.

To explain the fact that the gap at the DC remains open above $T_\mathrm{N}$ with preservation of its magnetic-derived origin we would like speculate proposing the following idea. Taking into account simultaneously a large SO coupling with preferential in-plane spin orientation of topological surface state and out-of-plane anisotropy of magnetic moments of Mn atoms characteristic for MnBi$_2$Te$_4$ one can assume a formation of spin fluctuations with a chiral spin texture similar to that described in Ref.~\cite{Liu2009} for magnetic impurities inside TI. Such a spin texture arises as a reaction of the medium (a topological insulator with enhanced spin-orbit coupling) to shield a local magnetic moment, which can appear on the corresponding magnetic defects or even on a PE-hole generated on a magnetic atom. As a result, the generated spin textures is characterized by a transformation of the spin orientation from the out-of-plane spin orientation in the center to the in-plane one under going to the periphery, like as it is for meron-like (or skyrmion-like) magnetic excitations, see, for instance, \cite{Lin2015,Jiang2019,Nagaosa2013}. We consider the formation of such excitations as a special kind of spin fluctuations developed in MnBi$_2$Te$_4$ above $T_\mathrm{N}$. They could predominantly induced, at least, around the magnetic interstitial defects, as shown in Ref.\cite{Jiang2019} or in the Mn-Te bilayers inside SL, like as it is in heterostructures MnTe/TI \cite{He2018}. The possibility of the skyrmion excitations in MnBi$_2$Te$_4$ and Mn-doped TI was experimentally confirmed in Refs.~\cite{Yan2019,Liu2017}. One can assume that laser excitation can additionally stimulate the generation of such meron-like excitations, as in Ref.~\cite{Ogawa2015}. These excitations (or spin fluctuations) play the role of emergent local space-modulated magnetic field, which is able to open the gap at the DC above $T_\mathrm{N}$.

 Fig.~\ref{ARPES2}, panels c1,d1 and panels c2,d2,  show the ARPES dispersion maps and the corresponding  EDCs for the sample with reduced gap, measured at  temperatures below and above $T_\mathrm{N}$, respectively. First of all, it can be clearly seen in Fig.~\ref{ARPES2} (c2 and d2) that the gap width does not depend on  polarization of LR both below and above $T_\mathrm{N}$. All presented EDCs  have a two-peak structure (similar to that shown in Fig.~\ref{ARPES1} (g)) regardless the temperature, that allows to conclude that this reduced gap also remains open above $T_\mathrm{N}$,  like in case of the large gap. It is interesting that both CD ARPES dispersion map (panel c3) and corresponding EDC (panel c4) measured for the Te-derived CB edge  states below $T_\mathrm{N}$ demonstrate   pronounced inversion of the CD signal. Although, the inversion of the sign for   upper and lower DCs   (panels c3 and c4) is practically not observed and can no longer be clearly traced in the CD measurements.   Whereas, some redistribution in the CD-signal between the upper and lower DC states can be distinguished in the CD ARPES dispersion maps. We associate this observation with a reduced effective magnetic moment developed for this kind of sample in the area of the topological DC state localization. At the same time, the CD-signal observed in the energy region of the exchange-split CB states demonstrates clear inversion of the sing between the split states. This splitting collapses above $T_\mathrm{N}$, that testifies rather to a developed FM magnetic coupling. It could means that despite surface FM magnetic ordering well probed by the exchange-split CB states, the DC states are affected by a significantly reduced effective out-of-plane magnetic moment. On the other hand, according to the theoretical modelling performed
in Ref. \cite{Hao2019} the ``gapless''-like  DC can arise in MnBi$_2$Te$_4$ due to the surface magnetic
reconstruction resulting in formation of a zero-like
out-of-plane magnetic moment in the Mn-layer inside the surface SL.
This may be due to (i) formation of the intralayer AFM coupling
(instead the FM one), (ii) in-plane arrangement of the magnetic
moments in the surface SL and (iii)  formation of the surface
spin-disordered paramagnetic-like magnetic layer. At the same time,
in these experiments a small energy gap at the DP of about 12-15~meV
can be also distinguished \cite{Hao2019,Li2019}.

However, our experimental spin-resolved and CD measurements demonstrate a rather well-developed surface FM ordering. At the same time, any visible $k_\|$ shift of the DC, which should occur in the case of the in-plane magnetization in a magnetic TI (see Refs. \cite{Shikin2018,Shikin2018a}) was also not observed in the measured ARPES dispersion maps for any LR polarizations and temperatures.

\noindent
\textbf{XMCD with varied applied out-of-plane magnetic field}

\begin{figure*}
\centering
\includegraphics[width=\textwidth]{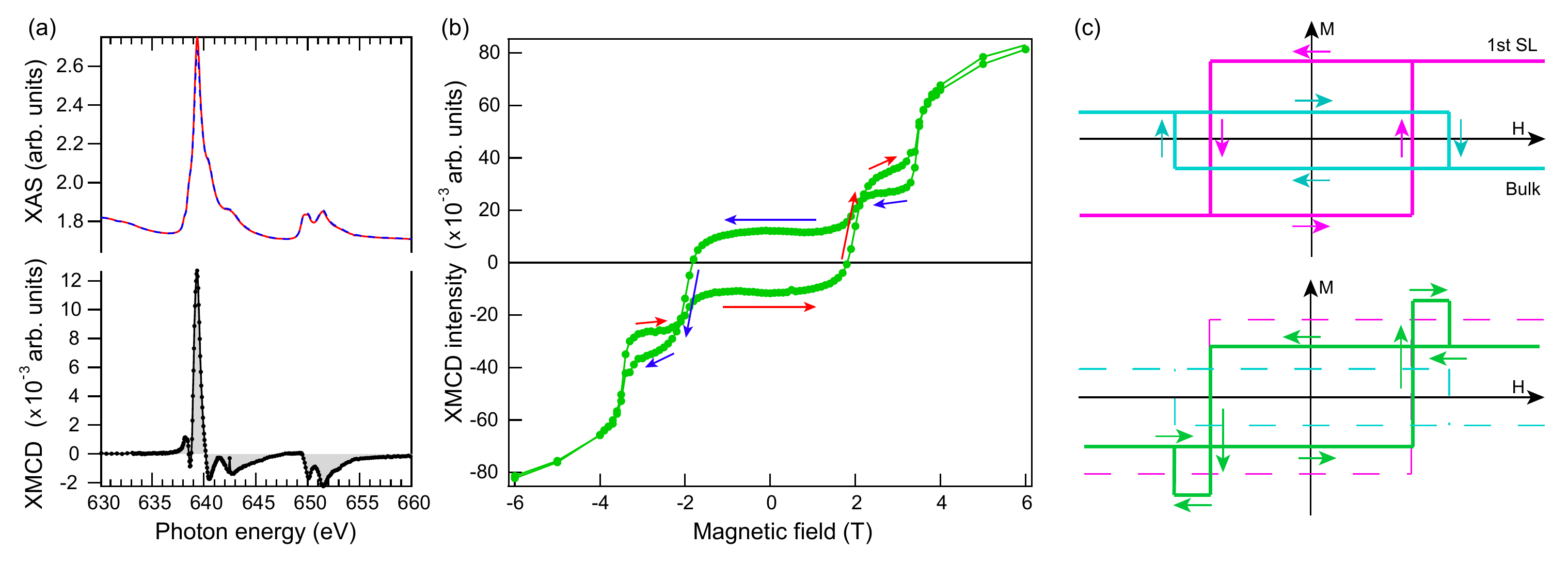}
\caption{ Dependence of the difference XMCD signal measured at the Mn $L_{23}$ absorption edge in total electron yield mode at temperature of 15~K for varied magnetic field (between -6 and +6~T) of opposite directions applied perpendicular to the surface. (a) - the Mn $L_{23}$ absorption edge (XAS) spectra measured at zero field with using opposite circular polarization of LR  and - corresponding difference between the spectra.  (b) - the variation of the resulting XMCD signal measured under applied magnetic field of opposite directions,  which is presented with the subtracted linear background.  (c) - schematic presentation of the hysteresis loops originated from the first and the second SL contributions with opposite magnetic moments (pink and blue curves) in formation of a total surface magnetization probed by XMCD and schematic presentation of the expected XMCD signal (green curve). }
 \label{XMCD}
\end{figure*}

To  study the surface magnetic ordering  we have carried out the XMCD measurement under variation of applied out-of-plane magnetic field of different sign. These measurements are more surface sensitive as compared to  the SQUID measurements. Fig.~\ref{XMCD} shows  XMCD results obtained with photon energy in the region 630-670~eV at temperature of 15~K for Mn $L_{2,3}$ photoexcitation edge. Fig.~\ref{XMCD} (a), upper part, shows the X-ray absorption spectra (XAS) measured at opposite circular polarizations. The XMCD signal, presented in the bottom part, was obtained as a result of subtraction of these spectra.  The XMCD signal is proportional to the induced magnetic moment under applied magnetic field. We note that the amplitude of the XMCD signal is comparable to that of the former report (see Fig.~2~(d) in Ref.~\cite{Otrokov2019}). Fig.~\ref{XMCD}~(b) shows the dependence of the resulting XMCD signal on variation of  sign and  magnitude of the applied perpendicular magnetic field. 

First of all, the presented XMCD measurements demonstrate a pronounced hysteresis loop observed in a weak applied magnetic field with an inverse sign of the induced surface magnetization for the opposite magnetic field. This observation allows us to conclude that the surface magnetic ordering is rather FM-like. Any features related to a possible surface AFM coupling along the surface in a weak magnetic field are not observed in the presented XMCD measurements. A stepped surface with terraces characterized by opposite spin orientations of the surface magnetic moments is also not able to demonstrate FM-like hysteresis loops presented in Fig.~\ref{XMCD}. 

At the same time, an unusual behavior of the hysteresis loop is observed in the region between 2 and 3.5 T where it demonstrates opposite directions relative to the ``main'' FM one. Appearance of these shifted hysteresis loops can be explained in simple approximation as follows. 
The XMCD signal (which is exponentially damped with a depth) is mainly provided by the contributions of the two upper surface SLs with opposite magnetic moments, with the reduced sensitivity to the second SL. These  contributions from two SLs and the resulting magnetic moment are shown schematically in Fig.~\ref{XMCD} (c). The arrows indicate the direction of the applied magnetic field. It is evident that the magnetic coercivity is different for these two opposite contributions. This can be related to inclusion of the surface topological DC state in the formation of the first SL magnetic texture. Bottom panel shows the resulting  averaged surface magnetic moment variation probed by XMCD in this case.

On the other hand, we have to note that such a complicated hysteresis structure is somewhat similar to the magnetic field-dependent Hall resistivity observed in Refs.~\cite{He2018,Liu2017,Yasuda2016}, where such behavior was associated with the formation of skyrmions. This is consistent with Ref.~\cite{Yan2019}, where the skyrmion formation in MnBi$_2$Te$_4$ was confirmed experimentally. 

Thus,  our magnetic measurements indicate an out-of-plane FM ordering, which has developed in the surface SL. At the same time, XMCD measurements demonstrate the importance of including the  contribution of the second SL with opposite magnetic moment to the averaged magnetic moment, which affects the  DC state.  As a result, the opening of the gap at the DP is rather managed by the effective reduced (compensated) magnetic moment developing in the region of the location of the DC states.

\noindent
\textbf{Structural effect on surface electronic structure of MnBi$_2$Te$_4$}

\begin{figure*}
\includegraphics[width=\textwidth]{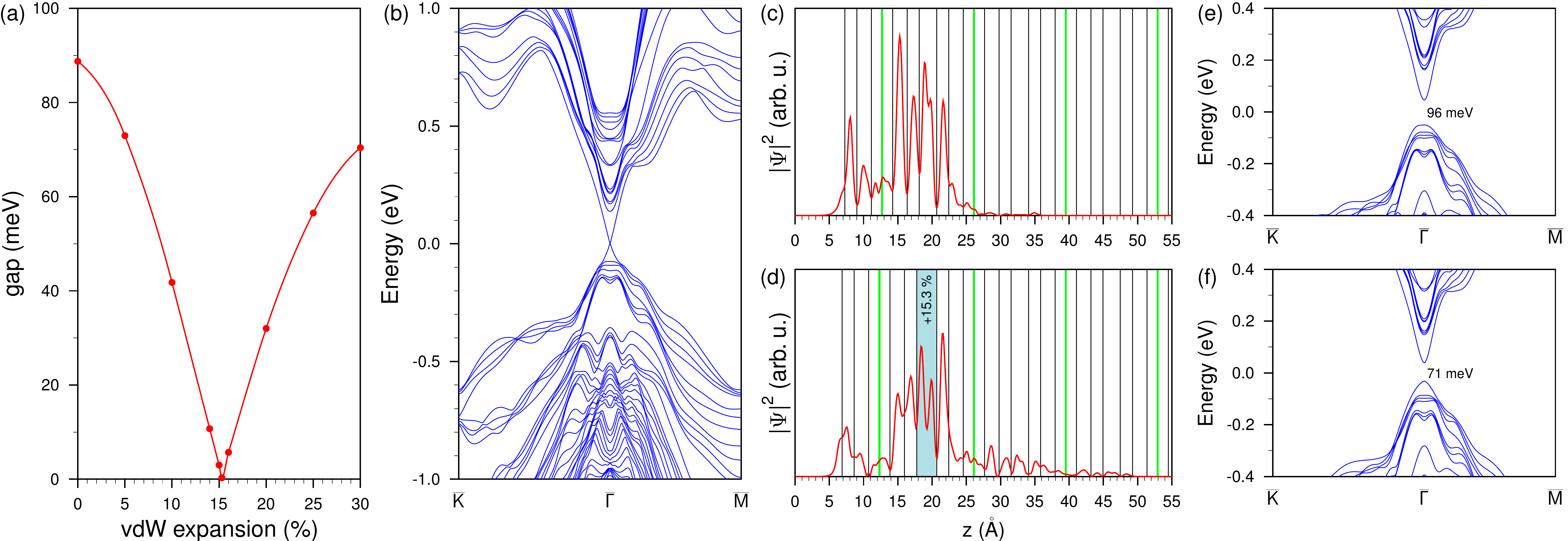}
 \caption{{\bf Dependence of the electronic structure of MnBi$_2$Te$_4$ on structure alteration.} (a) The gap in the Dirac state as function of the first van der Waals spacing expansion; (b) Electronic structure for the slab with the 1-st vdW spacing detached by 15.3~\%; Spacial charge distribution of the Dirac state at equilibrium structure (c) and for vdW spacing expanded by 15.3~\% (d); Surface electronic spectrum of MnBi$_2$Te$_4$ with modified interlayer distances in outer SL only (e) and outer SL and 1-st vdW spacing (f) as taken from Ref.~\cite{Zeugner2019}.}
 \label{fig_calc}
\end{figure*}

The alteration of the topological surface state localization is possible
due to structure modification caused by natural unavoidable defects.
It was revealed that MnBi$_2$Te$_4$ usually contains from 3~\% to
17.5~\% of Mn$_{\rm Bi}$ and Bi$_{\rm Mn}$ antisite defects as well
as Mn vacancies \cite{Liang2020,Yan_PRM2019,Zeugner2019}. Due to the
fact that Mn and Bi atoms have noticeably different size it is
evident that presence of a number of Bi$_{\rm Mn}$ defects in the Mn
layer should expand average interlayer Te-Mn-Te distances in the
middle of SL, and vice versa, the presence of Mn$_{\rm Bi}$ in the
Bi layers should lead to a decrease in the Bi-Te average interlayer
distances. Indeed, the structural parameters presented in
Refs.~\cite{Yan_PRM2019,Zeugner2019} show that Mn-Te and outer Te-Bi
interlayer spacings are respectively by 3--3.5~\% expanded and
contracted as compared to our calculated equilibrium interlayer
distances whereas the second, Bi-Te, distance differs only within
one percent from the theoretical value. At the same time these data
for defect containing samples also show that vdW spacing between
neighboring SLs is of $\sim$8-10~\% larger with respect to the
calculated MnBi$_2$Te$_4$ structure. Another structural effect
earlier discussed \cite{xu2012dirac} for TIs with weak van der Waals
coupling between building multilayered blocks is a broadening of vdW
spacing near the surface caused by the mechanical cleavage or
exfoliation implied for the surface preparation for ARPES and STM
experiments. For \emph{ab-initio} simulation of the structural
effects we apply an approach that previously showed its efficiency
in explaining the experimentally observed features in layered TIs
spectra and which is based on consideration of changes in
interplayer spacings within or between building blocks of TI
\cite{Eremeev_2012,Landolt_PRL2014,Pacile_RRL2018}. In our
simulation we restrict the structural changes only in the surface SL
and in the first vdW gap, since it is known that the topological
surface state is almost completely localized in this area. First we
consider the effect of vdW spacing expansion on the surface
electronic structure. As previously shown for layered TIs the
gradual detachment of the outer block leads to progressive
relocation of the topological state to the deeper QL
\cite{Eremeev_2012}. For MnBi$_2$Te$_4$ we have considered the vdW
spacing expansion up to 30~\%, first, in increments of 5~\% and
then, to find a minimum DP gap, with a smaller steps. As one can see in
Fig.~\ref{fig_calc} (a), the detachment of the outermost SL leads to
rapid decrease in the DC gap and at experimentally determined values
of expansion of $\sim$8-10~\% it is twice as smaller than in the
equilibrium structure. An intriguing finding is that at 15.3~\%
expansion of the first vdW spacing the topological surface state of
MnBi$_2$Te$_4$ AFMTI becomes gapless. The spectrum of the
MnBi$_2$Te$_4$ slab with outermost vdW spacing expanded by 15.3~\%
(corresponding to 0.38~\AA) is presented in Fig.~\ref{fig_calc} (b).
With a further increase in the outermost vdW spacing, the DC gap
rapidly increases. The strong dependence of the DC gap on the vdW
spacing variation stems from the change in spatial localization of
the topological surface state. Like in non-magnetic TIs, being
initially localized mainly in the surface SL
(Fig.~\ref{fig_calc} (c)) where it is affected by magnetization from
single Mn layer, it relocates inward and, as a result, begins to experience the magnetization provided by Mn atoms of the second SL
which are characterized by opposite orientation of the magnetic
moment. At certain expansion (which we identified is of 15.3~\%) the
influence of opposite magnetic moments compensate each other
(Fig.~\ref{fig_calc} (d)). With further increase in the outermost
vdW spacing the weight of the DC state in outer SL decreases even
more and the influence of the Mn layer of the second SL becomes
dominant. 

To study the effect of the structural changes produced by
antisite defects within SL we extracted the interlayer distances
from the structural data presented by Zeugner et al.
\cite{Zeugner2019}. The spectrum of the MnBi$_2$Te$_4$ slab with
modified in accordance with the experimental interlayer distances
outermost SL is shown in Fig.~\ref{fig_calc} (e). It demonstrates
the DC gap (96 meV) a bit wider than that in the equilibrium
structure (88 meV). After a series of calculations with a variation
in the SL interlayer distances we found that the gap can be smaller
than that in ideal MnBi$_2$Te$_4$ due to reduction of the second,
Bi-Te, spacing, but again not by much (at moderate variations).
Finally, introducing additionally experimental vdW spacing into the
structure, which is by 8~\% larger than theoretical spacing, we
expectedly got the gap reduction (71 meV, see Fig.~\ref{fig_calc}
(f)). Thus, considering two structural effects, intrablock effect,
provided by antisite defects within the SL block, and interblock vdW
spacing expansion, provided by both antisite defects and mechanical
cleavage distortions, we revealed that the former one does not
affect much the DC gap, whereas the expansion of the vdW spacing can
lead to wide variation in the gap width: from 88 meV in the ideal
structure to zero and hence the DC gap should depend strongly on a
sample quality and surface preparation accuracy. The gap may even slightly exceed the theoretical value for the ideal structure due to antisite defects inside the SL.

In the case of shifting the topological surface state location towards the second SL, the gap  at the DC can already be determined by interaction between the magnetic excitations (fluctuations) developed in the first and the second SL with opposite spin orientation and chirality (like bi-meron) \cite{Witten1979,Rosenberg2010,Nogueira2016,Swingle2011,Nogueira2018}. This many-body interaction can lead to effective fractionalization of
the axion term $\theta$ and corresponding discrete modulation of the gap 
at the DP. In such a case $\theta=\pi/4$ can be expected
\cite{Nogueira2016,Swingle2011,Nogueira2018} in comparison with
$\theta=\pi$ characteristic for ``ideal'' AFM MnBi$_2$Te$_4$ that
correlates with the difference in the experimentally observed large
and small gap values.

\section*{Conclusion}

We have shown that both a  large (62-67 meV) and
significantly reduced (15-18 meV) gap  at the DP can be
clearly distinguished using laser ARPES in the measured dispersion
maps for different kinds of samples (or different surface areas) of
the intrinsic AFM topological insulator MnBi$_2$Te$_4$. The first
value is close to the predicted from theoretical calculation. The
second one is close to the ``gapless''-like dispersion assumed for
the samples with significantly rearranged surface magnetic ordering.
In both cases the gap at the DP remains open above $T_\mathrm{N}$,
when the long-range magnetic ordering is destroyed, being only slightly
decreased in magnitude. A precise analysis of the gap width
as a function of temperature across the AFM transition (under
maintaining other experimental conditions) confirms that the gap
remains open under crossing the temperature of long-range magnetic
ordering. The magnetic transition is indicated in the ARPES
dispersion maps by a continuous decrease in the intensity of the DC gap
state  with temperature below $T_\mathrm{N}$. Above
$T_\mathrm{N}$, the intensity of the DC state is almost constant.

The measured spin-resolved ARPES dispersions demonstrate a helical
in-plane spin structure with inversion of the spin for the DC states
with opposite momenta both below and above $T_\mathrm{N}$. The
out-of-plane spin structure exhibits an inversion of the spin
polarization between the upper and lower DCs,
which is combined with the contribution determined by the PE final
state effect oscillating with photon energy. This in-plane and
out-of-plane spin distribution also  takes place above $T_\mathrm{N}$,
that may indicate a magnetic-derived origin of the gap opening,
including above $T_\mathrm{N}$. We propose the idea that above
$T_\mathrm{N}$ the gap at the DC can remains open by an emerging
short-range magnetic field generated by chiral spin fluctuations
(like vortex-skyrmions), which couple the DC state with opposite
momentum and spin orientation.

A detailed comparison of results obtained by
CD and XMCD measurements with  varied magnetic field for the samples with large and reduced gap confirms the surface
out-of-plane FM orientation, regardless of the gap value. However,
for the topological DC state, the CD ARPES measurements show a
different behavior: the CD signal demonstrates the sign inversion for the large gap samples and does not show it for the samples with the small (reduced) gap.  In addition to previously published
 explanations for the gap reduction, we have
shown by means of \emph{ab-initio} simulations that structural
effect consisting in strong dependence of the DC gap on the width of
vdW spacing, which is affected by antisite defects and perhaps by
mechanical cleavage distortions during surface preparation, can vary
the gap width in a wide range. This effect leads to inward shift of
the DC state localization where it becomes influenced by Mn atoms of
both first and second SLs, bearing opposite magnetic moments. The
simulation revealed that it even can results in zero gap at fairly
moderate value of vdW spacing expansion as of 0.38~\AA.  On the other hand, a
decrease in the gap can be associated with the many-body
fractionation effects developed under coupling between the chiral
spin fluctuations characterized by opposite magnetic moments
generated in the first and the second SLs. These assumptions require
further careful research.

\section*{Methods}

The measurements of the ARPES dispersion maps were carried at the
$\mu$-Laser ARPES system at HiSOR (Hiroshima, Japan) with improved
angle and energy resolution and a high space resolution of the laser
beam (spot diameter around 5~$\mu$) using a Scienta R4000 analyzer
with an incidence angle of the LR of 50$^\circ$ relative to the
surface normal \cite{Iwasawa2017}. For the experiment we used
photons ($h\nu$=6.3~eV) of different polarizations (p-, s- and
opposite circular ones).

The spin-resolved ARPES experiments were carried out using  s-
polarized LR ($h\nu$=7~eV) \cite{Yaji2019}  at the ISSP, University
of Tokyo.

The XMCD/XAS experiments were performed at the twin helical
undulator beamline BL23SU of SPring-8 \cite{Saitoh2012} in Japan.

High-quality MnBi$_2$Te$_4$ single crystals were synthesized using
the vertical Bridgman method and characterized by X-ray diffraction
\cite{Aliev2019} at the Institute of Catalysis and Inorganic
Chemistry of Azerbaijan National Academy of Science.

Part of the ARPES experiments were also carried out in the resource
center ``Physical methods of surface investigation'' (PMSI) at the
Research park of Saint Petersburg State University.

Clean surfaces of the samples were obtained by a cleavage in
ultrahigh vacuum. The base pressure during all photoemission
experiments was better that $1 \times 10^{-10}$~mbar.

\emph{Ab-initio} electronic structure calculations were carried
out within the density functional theory using the projector
augmented-wave (PAW) method \cite{PAW1,PAW2} as implemented in the
VASP code \cite{VASP1,VASP2}. The exchange-correlation energy was
treated using the generalized gradient approximation \cite{perdew}.
The Hamiltonian contained scalar relativistic corrections and the
Spin-orbit interaction was included in all types of calculations. In
order to describe the van der Waals interactions we made use of the
DFT-D3 \cite{Grimme} approach. The Mn $3d$-states were treated
employing the GGA$+U$ approach \cite{Anisimov1991} within the
Dudarev scheme \cite{Dudarev1998}. The $U_\text{eff}=U-J$ value for
the Mn 3$d$-states was chosen to be equal to $5.34\,\mathrm{eV}$, as
in previous works on MnBi$_2$Te$_4$
\cite{Otrokov.jetpl2017,Otrokov.2dmat2017,Eremeev.jac2017,EremeevNL2018,Otrokov.prl2019,Otrokov2019}.

\section*{Acknowledgements}

The authors acknowledge support by the Saint Petersburg State
University (Grant No. 40990069)), Russian Science Foundation (Grant
No. 18-12-00062 in part of the photoemission measurements and
Grant No. 18-12-00169 in part of the electronic band structure
calculations) and by Russian Foundation of Basic Researches (Grant
No. 18-52-06009) and Science Development Foundation under the
President of the Republic of Azerbaijan (Grant No. EI
F-BGM-4-RFTF1/2017-21/04/1-M-02). A. Kimura was financially
supported by KAKENHI (Grants No. 17H06138, No. 17H06152, and No.
18H03683). S.V.E. and E.V.C. acknowledge support by the
Fundamental Research Program of the State Academies of Sciences
(line of research III.23.2.9). The authors kindly acknowledge the
HiSOR staff and A. Harasawa at ISSP for technical support and help
with the experiment. The ARPES measurements at HiSOR were performed
with the approval of the Proposal Assessing Committee (Proposal
Numbers: 18BG027 and 19AG048). XAS and XMCD measurements were performed at BL23SU of SPring-8 (Proposal Nos. 2018A3842 and 2018B3842) under the Shared Use Program of JAEA Facilities (Proposal Nos. 2018A-E25 and 2018B-E25) with the approval of Nanotechnology Platform project supported by MEXT, Japan.

\section*{Authors contributions}

The manuscript was written by A.M.Sh. with active participation of  S.V.E., D.A.E. and I.I.K. All co-authors took part in the discussion and analysis of the experimental results. Preparation of the manuscript for publication with the presentation of the figures was carried out by D.A.E. ARPES measurements with use of LR including CD ARPES measurements were carried out by D.A.E, A.M.Sh., I.I.K., S.F, E.F.S., S.K. and A.K. Resonant photoemission experiments were performed by D.A.E, A.M.Sh., S.K and E.F.S.  D.A.E, A.M.Sh. I.I.K., S.F, K.M., T.O., K.K., K.Y. and S.Sh. took active part in measurements of spin-resolved PE spectra under excitation by LR and SR. XMCD measurements were carried out by I.I.K., D.A.E. A.K., Y.T., Y.S. Crystals were grown with following testing their crystalline structure by Z.S.A., M.B.B., I.R.A. and N.T.M. The band structure calculations were carried out by S.V.E., M.M.O. and E.V.Ch. The project was planned by A.M.Sh., D.A.E., I.I.K. and E.V.Ch.

\section*{Competing interests}
The authors declare that they have no competing interests.

\section*{Additional information}

The authors declare that the data supporting the findings of this study are available within the paper.

\bibliography{biblioGapPaper}

\end{document}